# Room-Temperature Ionic Liquids Meet Bio-Membranes: the State-of-the-Art

Antonio Benedetto[1,2]*

[1]School of Physics, University College Dublin, Dublin 4, Ireland
[2]Laboratory for Neutron Scattering, Paul Scherrer Institut, Villigen, Switzerland

*Corresponding author: antonio.benedetto@ucd.ie

ORCID: 0000-0002-9324-8595

**Abstract:**
Room-temperature ionic liquids (RTIL) are a new class of organic salts whose melting temperature falls below the conventional limit of 100°C. Their low vapor pressure, moreover, has made these ionic compounds the solvents of choice of the so-called *green chemistry*. For these and other peculiar characteristics, they are increasingly used in industrial applications. However, studies of their interaction with living organisms have highlighted mild to severe health hazards. Since their cytotoxicity shows a positive correlation with their lipo-philicity, several chemical-physical studies of their interaction with biomembranes have been carried out in the last few years, aiming to identify the microscopic mechanisms behind their toxicity. Cation chain length and anion nature have been seen to affect the lipo-philicity and, in turn, the toxicity of RTILs. The emerging picture, however, raises new questions, points to the need to assess toxicity on a case-by-case basis, but also suggests a potential positive role of RTILs in pharmacology, bio-medicine, and, more in general, bio-nano-technology. Here, we review this new subject of research, and comment on the future and the potential importance of this new field of study.

Key-words: Ionic liquids; biomembranes; phospholipid bilayers; toxicity; bio-medicine; nanotechnology.

**Introduction:**
Room-temperature ionic liquids (RTIL) are a vast class of ionic systems, usually consisting of an organic cation and either an organic or inorganic anion (Fig. 1), whose melting temperature falls below the conventional limit of 100 °C (Welton 1999). They have been intensively investigated for their potential applications as solvents, non-aqueous electrolytes, high-performance lubricants, and advanced engineering materials (Plechkova et al. 2009; Ranke et al. 2007b; Ghandi 2014). The widespread appeal of RTILs to some extent relies on their perceived low environmental impact, making these compounds one of the bases of the so-called "green-chemistry". Their introduction in industrial processes together with their organic character, in turn, motivated the first studies of their interaction with biomolecules, and bio-organisms (Petkovic et al. 2011). As a result, several studies have highlighted their toxicity on living organisms (Pretti 2006 ; Bernot et al. 2005; Ranke et al. 2006, 2007a; Stolte et al. 2007; Kulacki and Lamberti 2008). Toxicity is also a measure of the high affinity between RTILs and bio-systems. This affinity, in turn, together with the extremely tunable chemistry of RTILs, is at the basis of the great future of RTILs in applications from pharmacology, to bio-medicine, and, broadly speaking, in bio-nano-technology (Stoimenovski et al. 2012; Hough et al. 2007). It has been already shown, for example, that RTILs are able to:
  (i) kill bacteria;
  (ii) extract, purify and even store DNA at ambient temperature;
  (iii) stabilize proteins and enzymes;
  (iv) either help or prevent protein amyloidogenesis, in same cases even turning amyloid fibres back to functional proteins;
  (v) penetrate, create pores, and destroy biomembranes;
  (vi) dissolve cellulose and other complex polysaccharides.

A general overview of the interaction between RTILs and several classes of biomolecules (e.g. proteins and peptides, mono- and poli-saccharides, nucleic acids, and biomembranes) has been presented in two mini-review published in 2016 (Benedetto et al. 2016a, b). The main aim of these studies has been to link the biological effects of RTILs – usually detected by bio-chemical approaches – to the microscopic mechanisms of interaction between RTILs and biomolecules, with the idea that only a complete understanding of their microscopic mechanisms can provide the basis to (a) synthetize *greener* RTILs for industrial applications, and (b) develop breakthrough applications in bio-nano-technology. Having this *big-picture* in mind, the interaction of RTILs with biomembranes is one of the most relevant topics. Since the first encounter of any foreign chemical species with a living cell is likely to occur at its protective biomembrane, this subject merges the two original aims of these investigations, i.e. assessing and reducing RTILs cytotoxicity, and developing bio-nano-technologies. For example, it has been shown and suggested that:
  (a) controlling the poration of biomembranes by RTILs could result in the development of new drug-delivering methods, with an apparent impact in pharmacology;
  (b) even more, since their chemistry can be finely tuned in such a way RTILs can be lethal to bacteria at the same doses that are harmful to eukaryotic cells, they can open new antibacterial strategies, something that is quite needed nowadays;

(c) finally, tuning their chemistry already gave RTILs able to kill cancer cells and leave healthy cells almost unaffected.

Moreover, the water-like environment in which biomembranes reside are rich in a variety of (inorganic) ions, that play a major role in promoting and regulating the biomembrane functions. The effect of simple ions such as $Na^+$, $K^+$, $Cl^-$, etc. has been studied extensively with experimental and computational methods (Berkowitz and Vaćha 2012; Pabst et al. 2007; Böckmann et al. 2003) as well as by empirical modeling (Aroti et al. 2007). It is natural, therefore, to turn the attention to RTILs, i.e. organic salts, whose complex structure and larger size provide many more ways to tune their interaction.

This mini-review is dedicated to recent results obtained in the study of the interaction between RTILs and biomembranes, with a special focus on their chemical-physical properties.

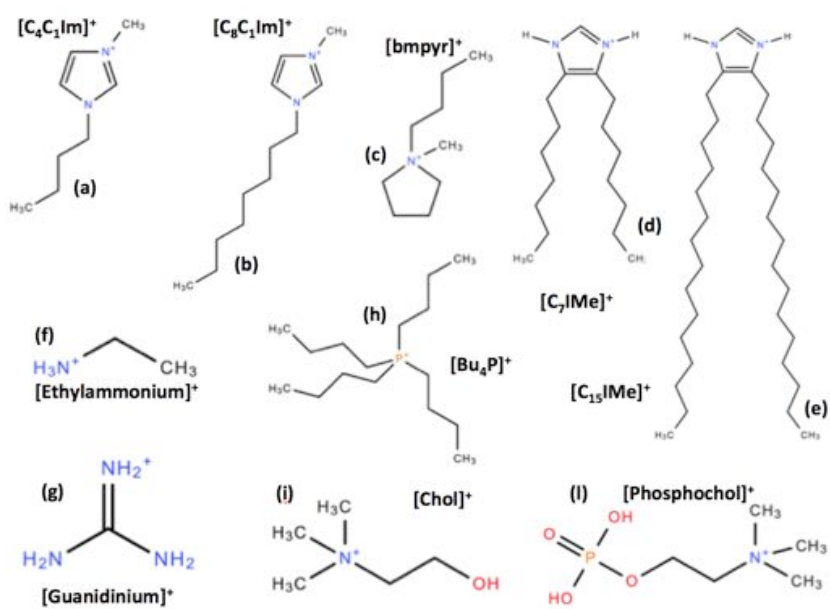

Fig. 1 – Chemical sketches of some selected RTILs cations: (a,b,c) the most common imidazolium and pyrridinium RTIL cations; (d,e) the double-tail lipid-mimic imidazolium-based RTILs of (Wang et al. 2015a); (f,g) the ethylammonium and guanidinium RTIL cations that help and contrast protein amyloidogenesis, respectively (Byrne 2007, 2008, 2009); (h) a phosphonium-based RTIL cation; and (i,l) the choline and phosphocholine cations also used in RTILs made of amino acids (Benedetto et al. 2014a).

**From complex biomembranes to phospholipids and their similarities with room-temperature ionic liquids:**

Biomembranes are complex and diverse biological supramolecular structures, which divide the inner part of cells where proteins, bio-complexes and bio-machineries, and nuclei reside, from the external environment. Moreover, they role in the biochemistry of cells is far more relevant than being just a physical barrier. Biomembranes, for example, regulate the diffusion of any chemical species into cells, either through specific (protein) channels or simply by absorption into the phospholipid region, and play a major role in cell replication processes as well. Biomembranes are also the target of several antibiotics, since even small modification of their structure, kinetics, and elastic properties can drastically affect the cells' stability up to their death. It is then not surprising that biomembranes are one of the major and quite broad subjects of studies. Here, we will review the small portion of those studies focused on the chemical-physical investigation of phospholipid bilayers. Phospholipid bilayers are a well-accepted first order model of biomembranes: they can be seen as the skeleton of any biomembranes into which proteins, extra lipids, saccharides, and, in general, bio-complexes can be absorbed to create more detailed and specific models of real biomembranes. As a result, assessing the effect of RTILs on phospholipid bilayers is the required first step along the path of the molecular-level comprehension of the biological effects of RTILs on biomembranes, and, in turn, on cells. This *modus operandi* is also corroborated by the fact that the cytotoxicity of RTILs measured by a variety of bioassay shows a clear positive correlation with their lipo-philicity (Ranke et al. 2006, 2007a; Stolte et al. 2007). Moreover, phospholipid bilayers can be easily prepared in a controlled and, thus, reproducible way in labs. Moreover, their cost is reasonable and affordable even for small research groups; this makes experimental investigations possible and accurate.

The basic units of any phospholipid bilayer are phospholipid molecules (Fig. 2a). They can be either zwitterionic or ionic, and are made by two hydrocarbon tails having a high hydrophobic character, and a hydrophilic head. As a result, when in a water environment, they arrange themselves in superstructures in such a way to minimize the contact between their hydrophobic tails and the water molecules, and maximize, on the other hand, the contact between their hydrophilic heads and the water molecules. Due to different factors, e.g. concentration of phospholipids in water, physico-chemical conditions, and geometry constrains, they can form uni- and multi-lamellar vesicles and micelles, and supported single-, bi- and multi-layers (Fig. 2b). Almost all of these geometries have been employed in the investigation of the effect of RTILs on biomembranes. Among all those systems, (i) supported bilayers, and (ii) vesicles are those that reproduce the double layer of biomembranes and, in turn, can be considered the best models to mimic them.

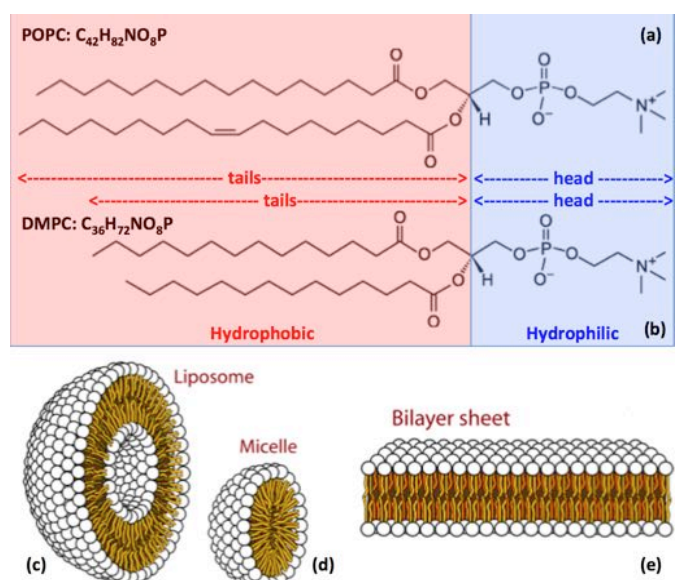

Fig. 2 – Two of the most common phospholipids: (a) POPC and (b) DMPC. They differentiate for the length of their (hydrophobic) hydrocarbon tails, whereas they share exactly the same (hydrophilic) head. When dispersed in aqueous environments, the hydrophobic-hydrophilic competition generates supramoleular structures like uni-lamellar (c) liposomes, (d) micelles, and (e) bilayer sheets (Wikipedia). Also multi-lamellar structures can be formed.

Comparing the phospholipids of Fig. 2a with the RTIL cations of Fig. 1 highlights their common similarities. Both types of molecules have an ionic / polar character, and both display hydrophilic and hydrophobic regions. The closest similarities is between phospholipids and the new RTILs synthetized by H.J. Galla and co-workers sketched in Fig. 1d (Wang et al. 2015a), which have two hydrocarbon chains as phospholipids. A second intriguing overlap is with the ionic liquids of amino-acids (AAIL) in which anions are deprotonated amino acids, and cations either a protonated choline or phosphocholine group that are also present in phospholipid head groups (Benedetto et al. 2014a).

In these two cases, but also in general, the similarities between phospholipids and RTILs are certainly responsible for their mutual affinity; moreover, they also suggest that a combination of (i) electrostatic, (ii) dispersion interactions, (iii) hydrophobic and hydrophilic effects, and (iv) hydrogen bonds structure and dynamics at the interface have to be taken into account to describe the mechanisms of interaction that perhaps should result to be a fine balance between all those forces in competition one to the other. It should then also be clear that even small changes in the chemistry of the molecules could affect the total balance of the forces, and, in turn, dramatically change the system properties. This circumstance highlights how a full comprehension of the microscopic mechanisms of RTILs –biomembranes interaction, together with the tunable character of the RTILs chemistry, is a key step for any major progress in this field. This observation is somehow in contrast with one of the main motivations at the basis of these investigations aiming at finding general rules to assess the effect of RTILs on biomembranes. However, we believe that this "unfortunate" circumstance can be balanced by the almost immense playground we have in front of us to develop new breakthrough applications in bio-nano-technology.

**A *snapshot* of RTILs - biomembranes interaction: a joint neutron scattering and computational study.**

Different techniques, both experimental and computational, have been used in the last decade to study the interaction between RTILs and (model) biomembranes, with the aim to determine the microscopic mechanisms behind the observed biological effects. Initial experimental studies carried out by K.O. Evans (Evans 2008) pointed to the marginal stability of phospholipid bilayers in contact with water solutions of imidazolium and pyrrolidinium RTILs. Both floating lipid vesicles and supported phospholipid bilayers geometries have been studied by means of photoluminescence, atomic force microscopy (AFM) and quartz microbalance, respectively. All measurements revealed damage to the bilayer from moderate to substantial, increasing with increasing length of the cation hydrocarbon tail. To the best of our knowledge, these are the first chemical-physics investigations of the interaction between RTILs and model biomembranes. However, the first experimental study able to characterize at molecular level the penetration of RTILs into model biomembranes has been done by means neutron reflectometry (Benedetto et al. 2014b). More specifically, by giving access to the density distribution of the chemical species (Fig. 3), neutron reflectometry has been used to investigate the changes in the microscopic structure and stability of two model phospholipid bilayers made by POPC and DMPC, respectively, in contact with water solutions of two RTILs ([bmim][Cl] and [Cho][Cl]). As a result it was observed that: (i) phospholipid bilayers maintain their characteristic 2D structure at the RTIL concentrations of the experiments (up to 0.5M); (ii) RTIL cations penetrate into the phospholipid region staying in the first leaflet at the junction between the phosphonium polar head and neutral hydrocarbon tail of phospholipid molecules (red curve in Fig. 3); (iii) the phospholipid bilayer thickness shrinks by about 1 Å, and the area per lipid increases; (iv) the amount of cations absorbed in the lipid region is more for DMPC than POPC; (v) the position of cations into the lipid region is independent of either the RTILs or the phospholipids choice; however (vi) for DMPC the [bmim]$^+$ cations have diffused in the inner layer as well. (vii) The penetration of the RTIL cations is not fully reversible, since after rinsing with pure water a non-negligible amount of cations remains in the lipid region (about 8% and 2.5% for DMPC and POPC, respectively).

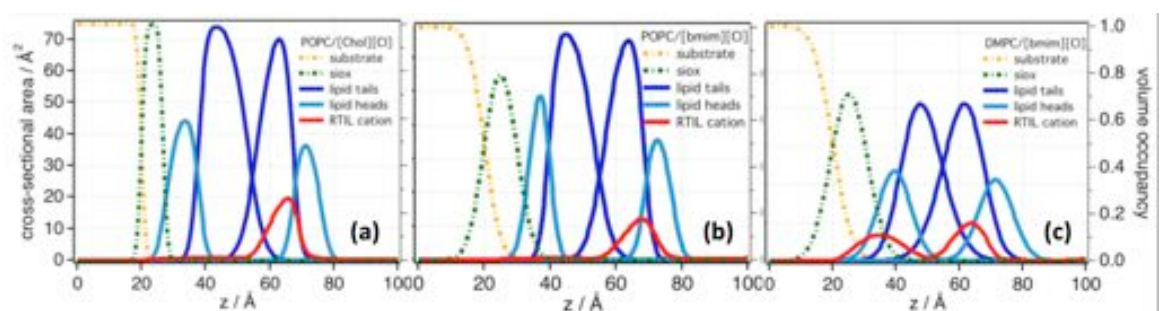

Fig. 3 – Density distribution profiles as a function of height $z$ from the surface of the substrate obtained by fitting the neutron reflectivity data taken from (Benedetto et al. 2014b). Neutron reflectometry has allowed to model each single supported phospholipid bilayers with 4 different density distributions accounting for (i) the inner lipid heads layer (cyan); (ii) the inner lipid tail layer (blue); (iii) the outer lipid tail layer (blue); and (iv) the outer lipid tail layer (cyan). (v) In red the density distribution of the cations, whereas the anion (Cl$^-$) is almost invisible to neutrons. Three cases are here reported where two different phospholipid bilayers interacr with water solutions of two different RTILs at 0.5M: (a) POPC and [Chol][Cl], (b) POPC and [bmim][Cl], and (c) DMPC and [bmim][Cl]; the RTILs cations absorption accounts for 8%, 6.5%, and 11% of the lipid bilayer volume, respectively. In (c) the diffusion of the cations into the inner leaflet is apparent. In all the cases phospholipid bilayers are in the liquid phase.

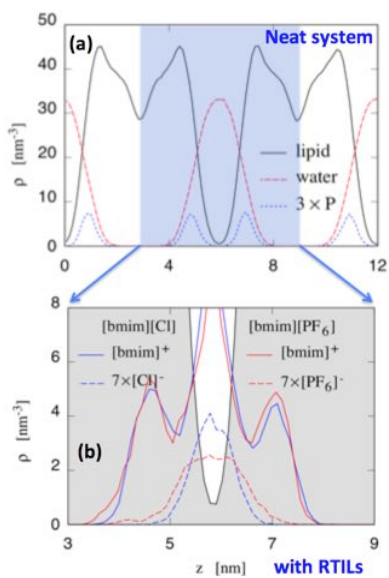

Fig. 4 - Density distribution profiles as a function of height *z* obtained from the full-atom classical MD trajectories of Ref. (Benedetto et al. 2015) for (a) neat POPC bilayers, and (b) doped with two RTILs. The computed profiles agreed with the one measured by neutron reflectivity reported in fig. 3: RTILs cations are absorbed in the lipid region whereas the anions remain in the water solution in contact with the bilayers.

Starting from the above-mentioned experimental results, A. Benedetto and co-workers carried out classical atomistic molecular dynamics (MD) simulations on the same systems (Benedetto et al. 2015). The MD simulations have reproduced the density distributions obtained by neutron reflectivity (Fig. 4), something that certifies the ability and good quality of the model. This is something that is not obvious and has to be checked, since the empirical force fields used in these MD studies, have never been tested properly for these ternary systems, i.e. phospholipids, water, and RTILs. Whereas the empirical potentials of lipids and water have been tested to work together, and the same is also true for those of RTILs and water, they never have been tested all three together. In a simplified way, we can conclude that the agreement between the MD simulations and the neutron reflectivity profiles (compare Fig. 3 with Fig. 4) is a proof of validity of the empirical potential used.

The MD simulations were able to determine or at least suggest the microscopic mechanism of the RTIL-biomembrane interaction, consisting of the following steps: (i) cations enter the phospholipid bilayer after 1-2ns of simulations; (ii) the penetration of cations into the bilayer is driven by the Coulombic attraction between the positive charge of the cation itself with the negative charged groups in the lipid head region (e.g. the negative oxygens in the carbonyl group at the matching point of the hydrocarbon tails), and (iii) it is stabilized by substantial dispersion forces between the cation and phospholipid tails; (iv) the absorption of the cations drives the penetration of a small amount of water into the polar portion of the phospholipid bilayer, and (v) stabilizes the hydrogen bonds at the lipid-water interface. Figure 5 presents some of the MD results.

In the MD study presented above, the systems were made of 1.360 POPC molecules, and 26.000 water molecules, and the computational "production" time was about 100-150 ns. The simulation box had a length of about 200 x 100 x 120 Angstrom along x, y, and z, respectively. For classical full-atom MD simulations, these are "big numbers", and to the best of our knowledge this is the biggest computer simulation study of phospholipids and RTILs. Both longer time scales and bigger systems are important to catch several features that otherwise cannot be detected. For example, only undulations of the bilayer surface whose wavelength is lower that twice the length of the simulation box can be detected, and only dynamical relaxations whose characteristic time is few times shorter that the total computational time can be properly identified. Needless to say, the simulation box sizes and computational "production" time affect also the error bars of all the observables extracted from the MD trajectories. As a result, our MD simulations allow to identify several trends due to the absorption of the RTIL cations in the lipid phase, from the shrink of the bilayer thickness, to the variation of diffusion coefficients and of elastic properties. We found, for example, that both the (i) isothermal (volume) compressibility, and the (ii) surface compressibility modulus increase upon the addition of RTIL, which also change the relaxation dynamics of hydration water and lipids by (iii) reducing the diffusion coefficient of water molecules, and (iv) increasing in [bmim][Cl] or

(v) decreasing in [bmim][BF$_6$] the diffusion coefficient of the lipids. Neutron scattering can be used to probe the most of these observables: elastic and quasi-elastic neutron scattering can access the pico-to-nanosecond dynamics (Benedetto and Kearley 2016; Magazu et al. 2010; Magazu et al. 2008; Bee 1988; Volino 1978); neutron spin-echo can be used to investigate slower relaxation processes (Mezei 1972), and also to probe structural fluctuations (Nagao 2009; Woodka 2012); small angle neutron scattering and diffraction can finally used for further structural characterization.

The joint experimental – simulation study presented above has its strength in combining two powerful approaches into one study, allowing a very detailed description of the microscopic mechanisms of RTIL-bilayer interaction. In the following we will present a series of results recently appeared in literature that either confirm or extend the scenario outlined above. Apart from few cases all the results are in agreement each other, and what is emerging is the extreme importance of the chemistry of lipids and RTILs, suggesting that case-by-case studies are needed. However, maintaining almost fixed the chemistry, some general trends emerge, like the correlation between RTILs chain lengths, concentration, and cytotoxicity.

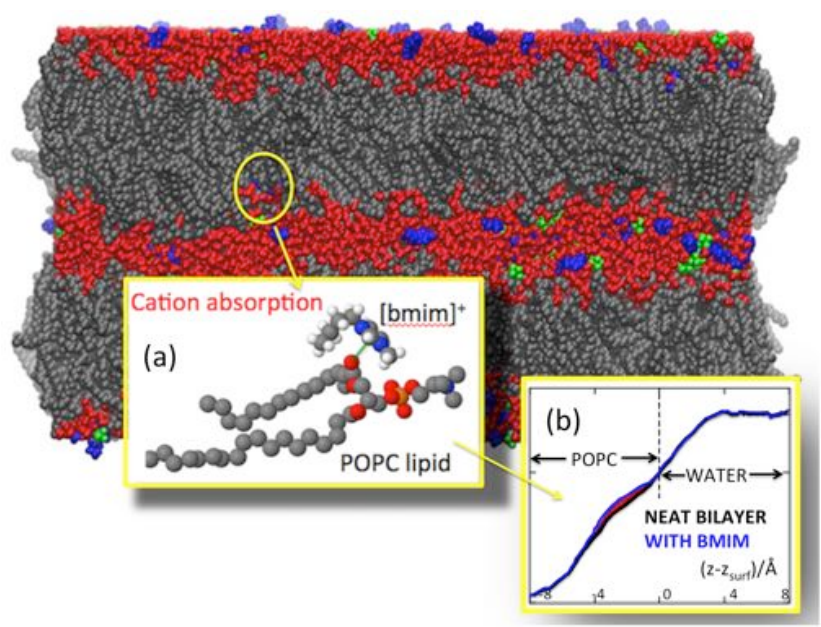

Fig. 5 - Schematic view of one of the sample configurations used in the MD simulations of (Benedetto et al. 2015). POPC domains in grey, water layers in red, [C$_4$mim]$^+$ in blue, and [PF$_6$]$^-$ in green. Inset (a): Representative configuration of POPC and [C$_4$mim]$^+$. Inset (b): water density profiles: the difference (area in red) points to a water excess in the POPC doped with [C$_4$mim]$^+$.

**What matters in the microscopic world of RTILs-biomembranes interaction? From chain length and RTIL concentration to chemistry: a huge playground of challenges and opportunities.**

The shrinkage of the phospholipid bilayer thickness upon the interaction with RTILs water solutions has been also confirmed by X-ray experiments (Bhattacharya et al. 2017, Kontro et al. 2016). In one case (Bhattacharya et al. 2017), single supported bilayers of DPPC (a zwitterionic lipid) interacting with water solutions of [bmim][BF$_4$] have been measured by means of X-ray reflectivity at several RTIL concentrations. As an additional result, the bilayer thickness seems to decrease faster with increasing RTIL concentration, although better error bars are needed to reach a final conclusion, see Table 1 of (Bhattacharya et al. 2017). The effect of concentration has also been studied by monitoring the survival percentage of *E. coli* versus the concentration of [bmim][BF$_4$], which shows an inverse correlation relationship (Fig. 6). In the same paper S.K. Ghosh and co-workers, report also that the in-plane elasticity of supported monolayers decreases upon addition of the RTIL, suggesting that the rigid structure of the well-packed lipid monolayer relaxes in the presence of the RTIL. The in-plane elasticity has been determined by the pressure-area isotherm profiles on monolayers, with the water solution of RTIL added to the monolayer-forming lipid solution. Another way to probe how the mechano-elastic properties are modified by the addition of RTILs, is offered by AFM, where both the Young's moduli and the rupture forces can be determined for single supported phospholipid bilayers and vesicles (Monocles et al. 2010; Garcia-Manyes and Sanz 2010; Roa et al. 2011; Ferenc et al. 2012; Stetter et al. 2016; Ding et al. 2017). To the best of our knowledge, there are not AFM studies of this kind so far. The elastic properties of phospholipid bilayers with and without RTILs have been studied by MD simulations as well; as discussed below. AFM can also image the surface of the bilayers and tell us something about the degree of homogeneity of the surface and/or the presence of pores and defects.

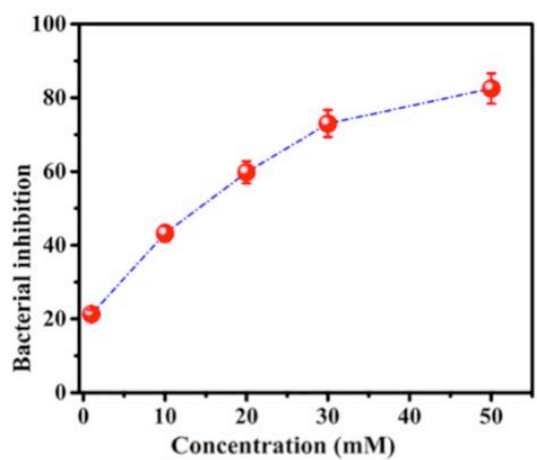

Fig. 6 – Inhibition percentage of *E. coli* versus the concentration of the RTIL [BMIM][BF$_4$] taken from (Bhattacharya et al. 2017).

The second X-ray study cited above (Kontro et al. 2016) used small-angle X-ray scattering (SAXS) to study multi-lamellar liposomes of eggPC and eggPG (80:20 mol%) – and also with cholesterol (60:20:20 mol%) – in interaction with water solutions of phosphonium-based RTILs. The results have been compared with those of the more common imidazolium-based RTILs. The interest of this study is increased by the fact that some phosphonium-based RTILs have shown antibacterial activity (O'Toole et al. 2012). The lamellar spacing of liposomes decreases with increasing RTIL concentration (Fig. 7). It is then clear that in this case RTILs pass through the phospholipids multilayer superstructures. In the same study S.K. Wiedmer and co-workers probe the effect of RTILs also by means of dynamic light scattering (DLS) and zeta potential measurements on large uni-lamellar vesicles. They conclude that the ability of RTILs to affect liposomes is related to the length of the hydrocarbon chains of their cations. They also conclude that the disruption of the phospholipid membrane is due to the disorder induced by the cation absorption. This result was confirmed by a more recent work again by S.K. Wiedmer and co-workers, where by means of nanoplasmonic sensing (NPS) measurement technique, they characterized the interaction between supported phospholipid uni-lamellar vesicles with amidinium- and phosphonium-based RTILs (Witos et al. 2017). NPS is a label-free optical technique that allows the study of surfaces and interfaces of metals, relying on surface plasmons, with a penetration depth of around 10 nm (quartz crystal microbalance has a penetration depth of about 250 nm).

S.K. Wiedmer and co-workers have also studied how the addition of cholesterol in the lipid phase changes the RTIL-biomembrane interaction (Kontro et al. 2016). This is a quite important, since it is well known that cholesterol is present in several biomembranes and has the ability to stabilize them. As a result, vesicles without cholesterol ruptured at lower concentrations than vesicles with cholesterol. At concentration the effects on the cholesterol-containing liposomes were more severe: the ruptured liposomes reassembled into organized lamellae, so under certain conditions, phosphonium-based ionic liquids have the ability to create new self-assembled structures from phospholipids.

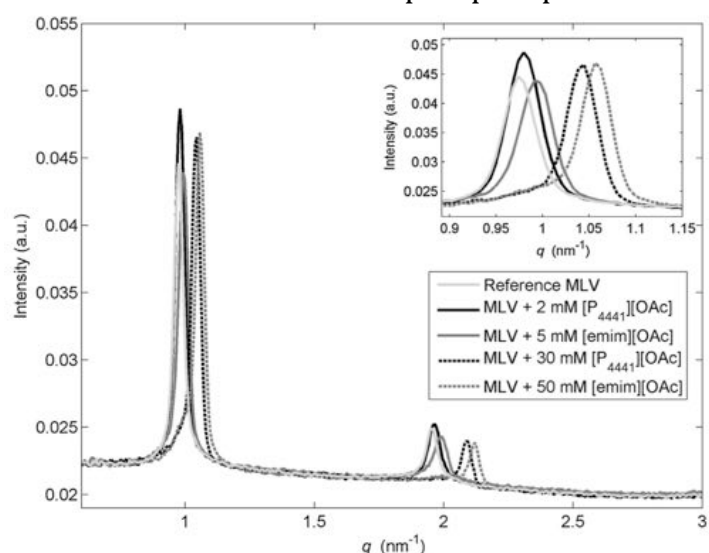

Fig. 7 – SAXS pattern of multilamellar POPC vesicles in interaction with RTILs from (Kontro et al. 2016). Reference MLV (light gray), MLV treated with [P4441][OAc] (black), and MLV treated with [emim][OAc] (dark gray). Insert: magnification of the first diffraction peak.

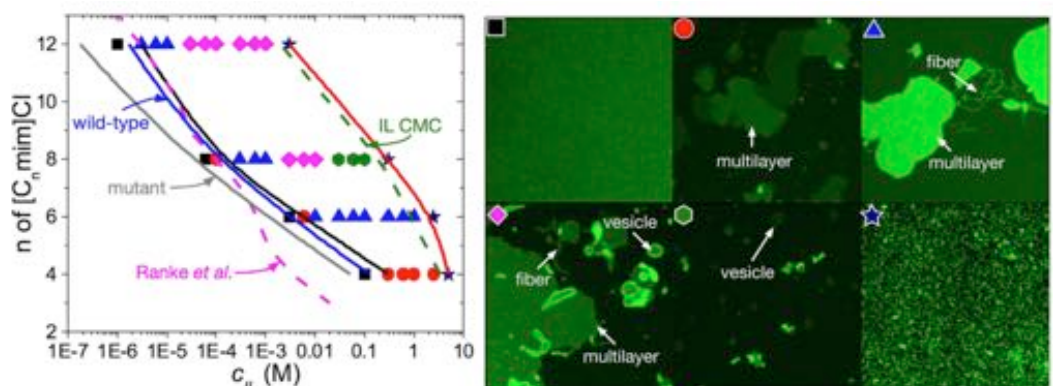

Fig. 8 – Phase diagram of [$C_n$mim][Cl] ionic liquid induced morphological changes to a supported α-PC bilayer taken from ref. (Yoo et al. 2016a). The EC50 toxicity line (in magenta) for IPC-cell shows a negative correlation between the toxic concentration and the RTIL cation chain length (Ranke et al. 2007a). The blue and lines are the predicted EC50 lines for wild-type (with cell wall)

and mutant (without cell wall) strains of *Chlamydomoas reinhardtii*, respectively. The green line corresponds to the RTIL critical micelle concentration (CMC) of (Blesic et al. 2007). The symbols correspond to the specific morphologies as in the right image. Black square: neat bilayer; red circle: multilayer; blue triangle: multilayer and fiber/tube; pink diamond: multilayer, fiber/tube and vesicle; green hexagon: vesicle; navy star: disrupted bilayer. The solid black and red lines correspond to the onset of supported lipid bilayer disruption and the total disruption of the supported lipid bilayer, respectively.

Further investigations on the connection between cytotoxicity, cation chain length, and RTIL concentration has been done by E.J. Maginn and co-workers by combining several experimental techniques with computer simulations (Yoo et al. 2014, 2016a, b). One of their main results is about the concentration-dependent effect of RTILs on biomembranes: they found that RTIL cations nucleate morphological defects on biomembranes at concentrations near the half maximal effective concentration (EC50) of several microorganisms, and that RTILs destroy biomembranes at the RTIL critical micelle concentration (Fig. 8). The results of E.J. Maginn and co-workers suggest that the molecular mechanism of RTIL cytotoxicity may be linked to the RTIL-induced morphological reorganization of cell membranes initially caused by the insertion of RTILs into the membrane. In their studies, they have also shown that cytotoxicity increases with increasing alkyl chain length of the cation (Fig. 8) and relate this observation to the higher ability of alkyl chain to penetrate, and ultimately disrupt, cell membranes.

E.J. Maginn and co-workers performed also full-atom and coarse-grained MD simulations. Coarse-grained simulations, in particular, by sacrificing some atomic-level details, are able to explore longer time-scales and larger length-scales than atomistic MD simulations; this is quite important to study both the absorption of RTILs, and the changes in the bilayers structures. Full-atom simulations usually can access a time scale of 100ns and a spatial scale of tens of nanometers, whereas with coarse-grained simulations it is possible to access time scales of microseconds and spatial scales of micrometers. As a result, their coarse-grained MD simulations show that the short-tail RTIL [$C_4$mim] cation spontaneously inserted into the lipid bilayer with the same orientation as that in the atomistic simulations, whereas the long-tail RTIL [$C_{10}$mim] cation self-assembled into micelle eventually absorbed onto the upper bilayer leaflet forming a RTIL monolayer. In the case of the short-tail cation, the number of inserted cations into the upper bilayer leaflet saturates at about 0.6 cations per lipid, after which the bilayer bends in response to the asymmetric distribution of inserted cations; moreover, while RTIL cations are absorbed, the bending modules of the bilayer drops from $22.6 \pm 1.7 \times 10^{-20}$ J to $9.3 \pm 0.9 \times 10^{-20}$ J. Since this asymmetric situation is not occurring when cations are absorbed in both the two leaflets, E.J. Maginn and co-workers comment on the inability of cations to diffuse from one leaflet to the other, that is at origin of the bending fluctuations of the bilayer (Fig. 9).

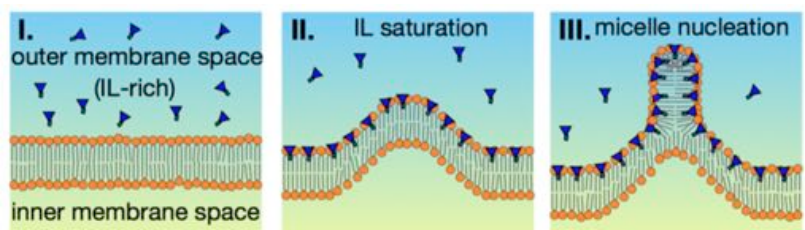

Fig. 9 – Coarse-grained MD simulations of Ref. (Yoo et al. 2016b) suggest that is the inability of some RTILs to diffuse from the outer leaflet to the inner leaflet of the phospholipid bilayer at the origin of the bilayer disruption.

The inability of the RTILs to diffuse into the whole bilayer after inserted into the closest leaflet seems in contrast, however, with the SAXS data of (Kontro et al. 2016), and in our opinion is something that requires more investigations. Having a closer look to our neutron reflectivity data of Fig. 3, for example, we can conclude that it seems to depend on the phospholipids and RTILs used. It seems, moreover, that it depends on the difference between the lipid and the cation chain lengths. Since [bmim][Cl] diffuses in the inner leaflet of DMPC (Fig. 3c), but does not with POPC (Fig. 3b), we could conclude that the shorter the lipid chain, the higher the ability of RTILs to diffuse from one leaflet to the other. Since E.J. Maginn and co-workers did not observe *inter-leaflet* diffusion of [bmim][Cl] in POPC, and S.K. Wiedmer and co-workers did observe *inter-leaflet* diffusion of different RTILs in POPC, we can conclude that the disagreement is due to the important role played by the RTIL chemistry, in line with our neutron reflectivity results (Benedetto et al. 2014b).

In our full-atom MD trajectories (Benedetto et al. 2015) done on POPC in interaction with [bmim][Cl], moreover, we did not observe any diffusion of cations from one layer to the other (result not yet published). However, longer simulation times may be needed to properly assess this *inter-leaflet* diffusion of RTILs, and its implication in RTILs cytitoxicity. AFM could also be a good technique to double-check indirectly this *in-bilayer* diffusion of RTILs by measuring the distance between the bilayer surface and the support: since we know that RTILs shrink the bilayer thickness, any increment of such distance could be related to the diffusion of RTILs into the inner bilayer leaflet and the water interlayer on the top of the support. The ability of RTILs to diffuse in multi-layer geometries can be useful not only in setting-up experiments (like diffraction experiments in which multi-layer lipid structures can be used) but also in applications. In our opinion this subject deserves further investigations.

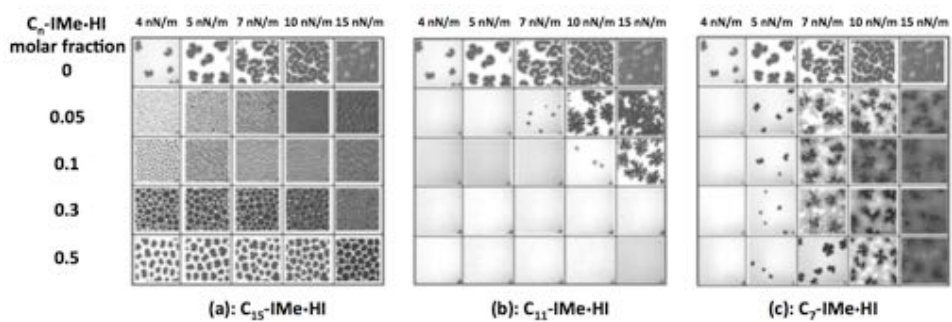

Fig. 10 – Epifluorescence images of mixed monolayers of DPPC with double-tail imidazolium-based RTILs of different chain-length at different molar fractions at the air-water interface at room temperature taken from (Wang et al. 2016). The membrane activity shows different

behaviors depending on the chain length of the RTILs: the bilayers (a) became more rigid for n=15, (b) get inhibited for n=11, but (c) almost unaffected for n=7. On the contrary, the biological activity of these RTILs goes the other way around, since the shortest one is the more toxic and the longest one stabilize the bilayer phase.

An important contribution to the saga has been made by H.J. Galla and co-workers (Wang et al. 2015a, b, 2016; Drucker et al. 2017, Ruhling et al. 2017). They have designed and synthesized a series of backbone-alkylated imidazolium-based RTILs composed of a hydrophilic N,N'-dimethylated imidazolium headgroup and two hydrophobic alkyl chains located at the 4- and 5-positions of the imidazole ring (Fig. 1d-e). The structure of these two-tail imidazolium-based RTILs ($C_n$IMe·HI) is similar to that of phospholipids (Fig. 2a-b), explaining the similarity of their physicochemical properties. The interest in this new class of RTILs is increased by their significant antitumor activity and cellular toxicity: in comparison with the more common one-tail imidazolium RTILs, they show approximately 3 order of magnitude higher antitumor activity. Having a global picture in mind, the most intriguing aspect is that their toxicity is negatively correlated with their chain lengths: the $C_7$IMe·HI has the highest toxicity and antitumor activity, whereas the $C_{15}$IMe·HI has the lowest toxicity. This circumstance is in contrast with the "general picture" presented above, and highlights how the chemistry of the molecules plays an extremely important rule. Moreover, the biological activity of these RTILs is inversely correlated with their lipo-philicity (Fig. 10), something that also plays against the general picture following which lipo-philicity and toxicity are directly correlated. From Fig. 10a it emerges that $C_{15}$IMe·HI has the higher membrane activity, and by increasing its concentration the RTIL-lipid system shows a reorganization at 0.3 molar fraction (more details in the referred paper). Fig. 10b shows that $C_{11}$IMe·HI has the ability to inhibit the supramolecular reorganization of the phospholipids, whereas fig. 10c shows how the membrane activity of $C_7$IMe·HI is negligible. These three different scenarios of RTIL-biomembrane interaction have been summarized in fig. 11.

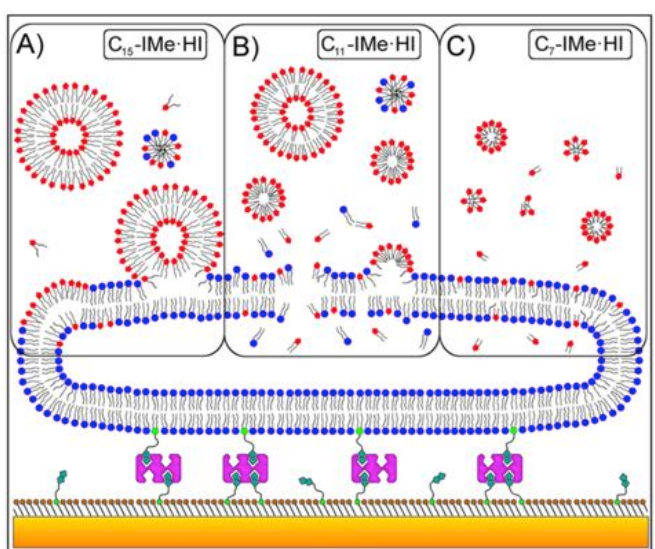

Fig. 11 – Model for membrane interaction and structure formation of double-tail imidazolium-based RTILs taken from (Drucker et al. 2017). Liposomes (blue) are tethered via biotin linkers (green) and streptavidin (purple) on a self-assembled monolayer (brown), which itself is chemisorbed on a gold-coated sensor surface (orange). (a) $C_{15}$-IMe·HI is able to form vesicles in solution that can then associate, fuse, and intercalate into bilayer membranes. (b) $C_{11}$-IMe·HI is able to form both vesicles and micelles while in water, which can then intercalate and lyse bilayer membranes. Bilayer disintegration is accompanied by the formation of micelles and mixed micelles. (c) $C_7$-IMe·HI dissolves to micelles and single molecules and can pass though the membrane without disintegration.

H.J. Galla and co-workers combined several experimental techniques as film balance, quartz crystal microbalance, confocal laser scanning microscopy, calorimetry, epifluorescence microscopic measurements, with MD simulations. They have also measured the interaction of RTILs with biomembranes enriched with cholesterol (Fig. 12).

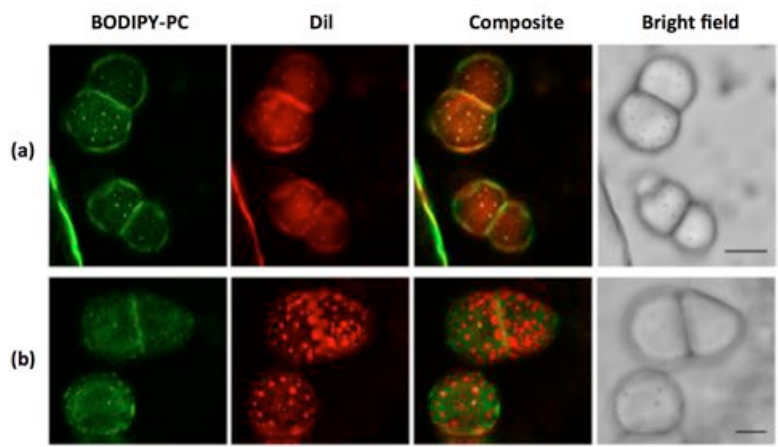

Fig. 12 – Bilayer domain fluidization of small bulged domains to flat large domains with enhanced dye specificity in the presence of 10% the double-tail imidazolium-based RTIL $C_{15}$IMe·HI from (Drucker et al. 2017). Giant uni-lamellar vesicles of (a) DOPC/SSM/Chol (33:33:33), and (b) DOPC/SSM/Chol/ $C_{15}$IMe·HI (33:23:33:10) at 38°C, scale 20 μm.

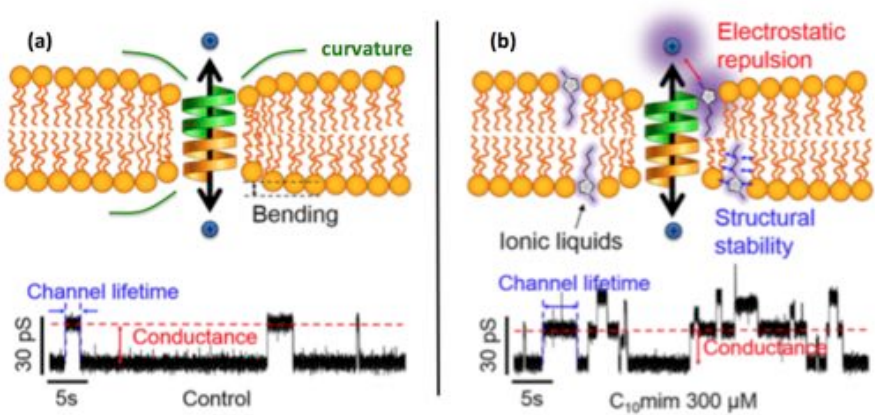

Fig. 13 – Effect of RTILs on gramicidin A ion channel from (Ryu et al. 2015). (a) Neat system; (b) system doped with RTIL. The RTIL cation $C_{10}$min (i) stabilizes the membrane-channel interaction by reducing the bilayer thickness and, in turn, its curvature closer to the channel location; and (ii) reduce the channel activity by electronic repulsion as sketched in (a). The function of the channel seems also affected by the amount of the inorganic salt NaCl in the solution: the higher the amount the higher the ion permeability.

Several other studies have been done on this new and promising subject, which more or less agreed with the results presented above: (i) positive correlation between RTILs chain lengths, RTILs concentration, and cytotoxicity (Witos et al. 2017; Losada-Pérez et al. 2016; Dusa et al. 2015; Galletti et al. 2015; Galluzzi et al. 2013; Kulacki and Lamberti 2007; Jeong et al. 2012; Mikkola et al. 2015; Jing et al. 2016); (ii) shrinkage of the bilayer thickness (Lim et al. 2014; Lim et al. 2015), and (iii) variation of its elasticity (Dusa et al. 2015) upon the absorption of RTIL cations; and (iv) importance of the chemistry of the molecules (Weaver et al. 2013; Gal et al. 2012; Lee et al. 2015). The effect of RTILs on the thermotropic behavior of biomembranes have also probed in several works (Weaver et al. 2013; Wang et al. 2016; Jeong et al. 2012): at low concentrations of RTILs the variation of the main phase transition of phospholipids bilayers is negligible, and at high concentration the variation is of the order of 5 to 10 degrees, sometimes reaching 20 degrees, but this just before the collapse of the bilayers. Several other MD simulations studies have also done on this subject (e.g. Bingham and Ballone 2012; Cromie et al. 2009; Lim et al. 2015), which give access to the detailed mechanisms of interactions. In Lim et al. 2015, for example, the mutual interaction between RTIL cations once absorbed into the membrane have been studied, and the measured increment of permeability due to the absorption of RTILs has been related to their antibacterial activity. Investigations focuses on specific cases and more complex systems, moreover, have enriched even more the broad and diverse panorama presented above (Patel et al. 2016; Modi et al. 2011; Jeong et al. 2012; Ryu et al. 2015; Lee et al. 2015). An interesting example is the study of T-J Jeon, H. Lee and co-workers on the effect of RTILs on the ion channel function of Gramicidin A embedded in a phospholipid bilayer (Jeong et al. 2012; Ryu et al. 2015; Lee et al. 2015). Their results show for the first time how the changes of the physical properties of the biomembrane (e.g. thickness) induced by the absorption of RTILs cations can influence the activities of membrane proteins (Fig. 13). These effects are more significant with RTILs with longer alkyl chains, and at higher RTILs concentration. Interestingly, they figured out how the concentration of inorganic salts (i.e. NaCl) can play a major role also in this case. The picture that emerges is the following: RTILs disorder phospholipids and shrink the bilayer, which yields less membrane curvature around the gramicidin A and thus stabilizes it, leading to the increased ion permeability. However, this effect occurs at 1 M of NaCl, where RTILs only slightly increase the phospholipids dynamics because of the strong electrostatic interactions between NaCl and lipids. On the other hand, at 0.15 M of NaCl, RTILs also significantly increase the lateral mobility of both phospholipids and gramicidin A, which leads to the decreased ion permeability.

**Summary and outlook for the future**
The analysis of the current literature on the interaction between RTILs and biomembranes shows that the first target of these studies has been to determine the microscopic mechanism behind the cytotoxicity of RTILs with the aim to design *greener* RTILs for industrial applications. Whereas a positive correlation exists between RTILs chain lengths, RTILs concentration and cytotoxicity, the are several exceptions that highlight how the RTILs effect on biomembranes is a complex balance of different interaction where the chemistry of each molecule is playing a non-negligible role. On the one hand, this observation implies that the effect of RTILs on biomembranes has to be assessed on a case-by-case basis. On the other hand, the same observation opens a huge playground of new opportunities for applications in bio-nano-technology. In our opinion these opportunities are more important than the difficulties preventing an *a priori* assessment of the potential danger. Just few nano-bio-technology studies have been done so far, and we are just at the beginning of this promising field of research.

**Acknowledgment:**
A.B. thanks Prof. Pietro Ballone for fruitful discussions. A.B. acknowledges support from (i) the European Community under the Marie-Curie Fellowship Grants HYDRA (No. 301463) and PSI-FELLOW (No. 290605), and from (ii) Science Foundation Ireland (SFI) under the Start Investigator Research Grant 15-SIRG-3538, with additional support provided by the School of Physics, University College Dublin, Ireland, and the Laboratory for Neutron Scattering, Paul Scherrer Institute (PSI), Switzerland.